# Design of a Plastic Inorganic Semiconductor GaPS$_4$-Based Gas Sensor for Conformal Monitoring of Gas Lines

*Qiao Wang[1,2]*

1. Key Laboratory of Micro-systems and Micro-structures, Manufacturing of Ministry of Education, Harbin Institute of Technology, Harbin 150001, China

2. School of Chemistry and Chemical Engineering, Harbin Institute of Technology, Harbin 150001, China

ABSTRACT

This paper reports the first gas sensor based on the plastic inorganic semiconductor $GaPS_4$, pioneering the application of plastic inorganic semiconductors in the field of gas sensing. Unlike traditional rigid sensors, this device leverages the unique layered structure and ultra-wide bandgap of $GaPS_4$ to achieve high sensitivity and selectivity in detecting $NO_2$. The intrinsic plastic deformability of the material enables it to conform tightly to complex curved pipelines like an "electronic bandage," completely eliminating monitoring blind spots. Nanoindentation tests reveal that its extremely low hardness (0.20 GPa) confers exceptional flexibility while maintaining stable electrical characteristics even under bent states. The device exhibits a linear response to $NO_2$ concentrations ranging from 1 to 10 ppm at room temperature. Although the limited defects in the single-crystal material result in pA-level response currents, defect engineering offers a viable pathway for performance enhancement. This study breaks through the conventional boundaries of plastic inorganic semiconductors confined to photoelectric and thermoelectric applications, opening new avenues for their use in gas sensing and advancing gas monitoring technology toward "conformal integration."

**1. Introduction**

The safety monitoring of industrial and civilian gas pipelines, chemical process lines, and other critical infrastructure is vital for ensuring production safety, preventing environmental pollution, and avoiding resource waste. Traditional gas monitoring technologies, especially for detecting pipeline leaks or process gases like highly corrosive and toxic nitrogen dioxide ($NO_2$), have long faced a series of core challenges: rigid sensors struggle to achieve conformal contact with pipes featuring complex curvatures, leading to monitoring blind spots and dead zones; their high power consumption limits long-term deployment in remote or distributed scenarios; simultaneously, achieving high selectivity, long-term stability, and mechanical robustness in complex gas backgrounds often remains difficult. Consequently, developing a novel gas sensing technology that can seamlessly integrate with pipeline surfaces while combining high performance with exceptional environmental adaptability has become an urgent goal in the field.

This paper presents the design of an innovative gas sensor based on the plastic inorganic semiconductor $GaPS_4$, designed to address these challenges and enable the conformal monitoring of gas lines. The core of this detector lies in utilizing a wide-bandgap p-type plastic inorganic semiconductor material, whose unique properties enable a revolutionary combination of performance characteristics.

On the performance front, the material's intrinsic properties form the foundation for its exceptional advantages: based on the strong electron extraction interaction between $NO_2$ and the p-type semiconductor, the detector exhibits high sensitivity and high selectivity towards $NO_2$. Its wide-bandgap nature further ensures low cross-sensitivity to reducing gases, water vapor, and environmental oxygen, while also endowing the device with low power consumption (capable of operating at room temperature) and excellent high-temperature stability, significantly outperforming organic semiconductor materials.

More revolutionary is the application breakthrough enabled by its plastic deformability. This detector can conform directly and tightly to curved, irregular, or vibrating pipeline surfaces, much like an "electronic bandage," achieving unprecedented conformal attachment. This not only eliminates monitoring blind spots but also enables accurate local concentration measurement at critical locations like valves and joints. Furthermore, the material's inherent plastic deformation mechanism grants it outstanding mechanical robustness and fatigue resistance, allowing it to withstand long-term deformation without failure. This characteristic paves the way for integrating high-performance sensors into wearable devices or inspection robot surfaces, enabling distributed, mobile smart monitoring networks.

The $GaPS_4$-based gas sensor designed in this work successfully merges high-performance gas sensing characteristics with customizable deformability. It signifies a paradigm shift in gas monitoring technology from simple "installation" towards intelligent "conformal integration" and "symbiosis" with the infrastructure, offering a highly promising novel solution for achieving ubiquitous, accurate, reliable, and long-term stable safety monitoring of gas lines.

## 2. Results and Discussion

Two-dimensional ultra-wide bandgap semiconductor $GaPS_4$ demonstrates significant potential in cutting-edge fields such as deep-ultraviolet photodetection, polarized optical sensing, and high-temperature neuromorphic computing, owing to its exceptional layered structure, tunable electronic properties, and outstanding environmental stability.[1–3] In this study, high-quality $GaPS_4$ crystals were successfully synthesized via chemical vapor transport (Figure S1). Specifically, stoichiometric mixtures of Ga, P, and S elemental powders in a 1:1:4 molar ratio were sealed in an evacuated quartz tube and subjected to a temperature gradient of 750/710°C, ultimately yielding layered bulk single crystals. X-ray diffraction analysis (Figure 1B) confirmed that the synthesized $GaPS_4$ crystals adopt a monoclinic structure with the space group $P2_1/c$. The diffraction peaks match well with the standard reference pattern (JCPDS No. 22-1476), indicating high crystallinity and a preferred orientation along the a-axis. Scanning electron microscopy (Figure S2A) coupled with energy-dispersive X-ray spectroscopy (Figure S2B) further verified the chemical homogeneity of the material, showing uniform distribution of Ga, P, and S elements with an atomic ratio close to the theoretical stoichiometry of 1:1:4. Structurally, GaPS$_4$ features a highly anisotropic two-dimensional covalent network within each layer (Figure 1A), formed by edge-sharing [GaS$_4$] and [PS$_4$] tetrahedra, while the layers are stacked via van der Waals interactions, with an interlayer spacing of approximately 2.74 Å and a single-layer thickness of about 0.69 nm. As an ultra-wide bandgap semiconductor, GaPS$_4$ exhibits a unique electronic structure: first-principles calculations predict a direct fundamental bandgap of 3.87 eV for the monolayer at the Γ point, whereas experimentally measured optical bandgaps are significantly larger, ranging from 4.22 to 4.50 eV. This discrepancy originates from the inversion symmetry inherent in the C$_{2h}$ point group of GaPS$_4$, which forbids dipole-allowed transitions between the valence band maximum and the conduction band minimum. As a result, optical absorption arises from higher-lying valence band states, leading to the observed blue-shifted absorption onset.

This unique layered structure, combined with its ultra-wide bandgap characteristics, makes GaPS$_4$ an ideal material for high-performance gas sensors: First, the layered structure provides a large specific surface area and abundant active sites, offering an ideal interface for gas molecule adsorption and reaction. Second, the ultra-wide bandgap ensures intrinsic stability under high-temperature and harsh environmental conditions, effectively suppressing thermally excited leakage currents and significantly reducing the device's baseline noise. Third, the van der Waals gaps between the two-dimensional layers provide controllable channels for the selective diffusion and intercalation of gas molecules, which is conducive to improving the sensor's response speed and selectivity. Fourth, the anisotropic charge distribution within the structure enhances orbital coupling and charge transfer efficiency between the material surface and specific gas molecules, laying the foundation for high-sensitivity detection. The synergistic effects of these structural and band characteristics endow GaPS$_4$ with unique application advantages in the field of high-temperature, high-sensitivity gas sensing.

GaPS$_4$ demonstrates mechanical behavior that challenges conventional understanding—this wide-bandgap semiconductor material exhibits metal-like plastic deformation capability. As shown in Figure 2A, a nearly 1 mm thick GaPS$_4$ single crystal can be bent into a U-shaped structure without fracture, a deformation scale extremely rare in inorganic single crystals. Microstructural analysis (Figure 2B) reveals that under bending conditions with a 40 μm radius of curvature, the material only experiences slight interlayer slip without any crack propagation, demonstrating its unique plastic deformation mechanism.

Nanoindentation test data unveil the essence of this anomalous mechanical behavior (Figure 2C). The hardness value of GaPS$_4$ is merely 0.20 GPa, which carries significant physical implications: it is two orders of magnitude lower than that of typical brittle material silicon (13.29 GPa) and even lower than the known plastic material GaTe (0.71 GPa). Particularly noteworthy is that GaPS$_4$'s hardness is only about one-ninth that of polycrystalline copper, indicating extremely low deformation resistance, which fundamentally explains why millimeter-scale crystals can achieve large-curvature bending.

This unique combination of mechanical properties brings revolutionary breakthroughs for a new generation of flexible gas detectors. Conventional gas sensing materials generally suffer from brittleness issues, making them difficult to conformally integrate onto irregular surfaces. In contrast, GaPS$_4$'s exceptional plasticity enables direct conformal attachment to various complex curved surfaces (such as key monitoring areas like gas-lines and valves, Figure3A). As demonstrated in Figure 3B, its extremely low hardness ensures tight contact with the measured surface, effectively avoiding detection blind spots caused by poor contact in traditional rigid sensors. Meanwhile, the material maintains stable layered structure and electronic characteristics even under bent states, providing a crucial material foundation for developing flexible, wearable high-performance gas detection systems, significantly expanding the application potential of gas sensors in industrial safety monitoring and environmental detection.

Building upon its exceptional mechanical properties, we further constructed a $GaPS_4$-based gas sensor and systematically evaluated its detection capability for $NO_2$. As shown in Figure 3C, the device demonstrates effective detection of $NO_2$ at concentrations as low as 1 ppm, with the response current exhibiting a favorable linear increase against the gas concentration ranging from 1 to 10 ppm (Figure 3D). It is worth noting that due to the use of high-quality single crystals with minimal internal defects and grain boundaries—which serve as active sites for gas adsorption— coupled with limited charge transfer efficiency between the material surface and gas molecules at low concentrations, the absolute response current remains at the pA level with a relatively small magnitude of increase. These findings indicate the potential of this device for quantitative detection of low-concentration $NO_2$, while also suggesting that introducing additional active sites through controlled defect engineering could be an effective strategy to further enhance its gas sensing response.

## 3. Conclusions

This study successfully designed and fabricated a novel gas sensor based on the plastic inorganic semiconductor $GaPS_4$. Leveraging its unique layered structure and ultra-wide bandgap properties, the device achieves high sensitivity and selectivity in detecting $NO_2$. The intrinsic plastic deformability of the material enables conformal attachment to complex curved pipelines, effectively addressing the monitoring blind spots inherent in traditional rigid sensors. Nanoindentation tests reveal that the extremely low hardness (0.20 GPa) of GaPS$_4$ endows it with exceptional flexibility, allowing it to maintain stable electrical characteristics even under bent states. In terms of gas sensing performance, the device exhibits a linear response to NO$_2$ concentrations as low as 1 ppm at room temperature, demonstrating its potential for quantitative low-concentration gas detection. Although the limited defects in the single-crystal material result in a pA-level response current, controlled defect engineering may further enhance its gas sensing performance. This work establishes a material foundation for a new generation of flexible and wearable gas sensing systems, signifying a paradigm shift in gas monitoring technology from simple "installation" to intelligent "conformal integration," with promising applications in industrial safety and environmental monitoring.

## 4. Experimental Methods

***Growth of single-crystal GaPS$_4$:***

High-purity Ga (5N, 99.999%), P (4N, 99.99%), and S (5N, 99.999%) were obtained from ZhongNuo Advanced Material (Beijing) Technology Co., Ltd., and used as raw materials.

Bulk GaPS$_4$ single crystals were synthesized via a chemical vapor transport (CVT) method. Briefly, a quartz ampoule was loaded with a stoichiometric mixture of gallium (Ga), phosphorus (P), and sulfur (S) powders in a molar ratio of 1:1:4, subsequently evacuated and sealed under vacuum. To account for the differing boiling points and vapor pressures of the constituent elements, an inclined reaction vessel was constructed using an internal quartz holder. The chemical vapor transport (CVT) process was conducted in a dual-zone furnace with a thermal gradient spanning from 750 °C (high-temperature zone) to 710 °C (low-temperature zone) over a distance of approximately 30 cm. This thermal profile was maintained for two weeks at a controlled heating rate of 1 °C min$^{-1}$. Thereafter, the temperatures of both zones were gradually lowered to allow natural cooling to room temperature, resulting in the formation of bulk GaPS$_4$ crystals.

***Characterization:*** The scanning electron microscope (SEM) images were carried out on ZEISS Merlin Compact. Optical photographs were taken by VIVO IQOO Z7I. The GaPS$_4$ crystal was examined by X-ray diffraction using Cu Ka Source (Panalytical Instruments, X 'PERT). Nanoindentation data were collected on an Agilent G200. The device test was recorded using a Keithley 2400.

**Notes**

The authors declare no competing financial interest.

REFERENCES

[1]    D. Cao, Y. Yan, M. Wang, G. Luo, J. Zhao, J. Zhi, **2024**, *2314649*, 1.

[2]    Y. Yan, J. Yang, J. Du, X. Zhang, Y. Liu, C. Xia, Z. Wei, **2021**, *2008761*, 1.

[3]    T. Shen, C. Zhang, C. Qiu, H. Deng, **2025**, DOI 10.1063/5.0089393.

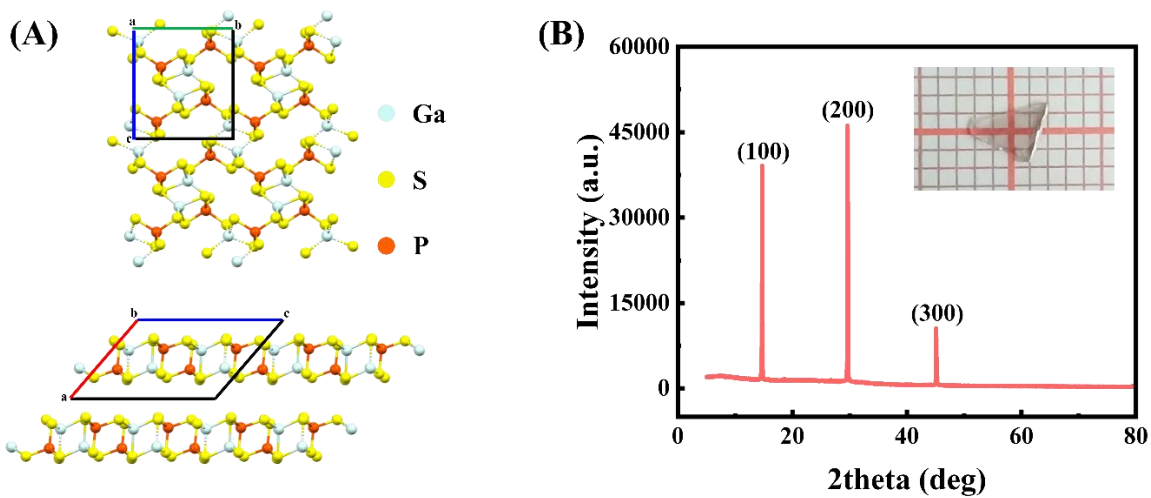

**Figure 1. Structural and Crystallographic Characterization of GaPS$_4$.** (A) Crystal structure of monoclinic and GaPS$_4$. (B) X-ray diffraction (XRD) spectra of a GaPS$_4$ single crystal flake.

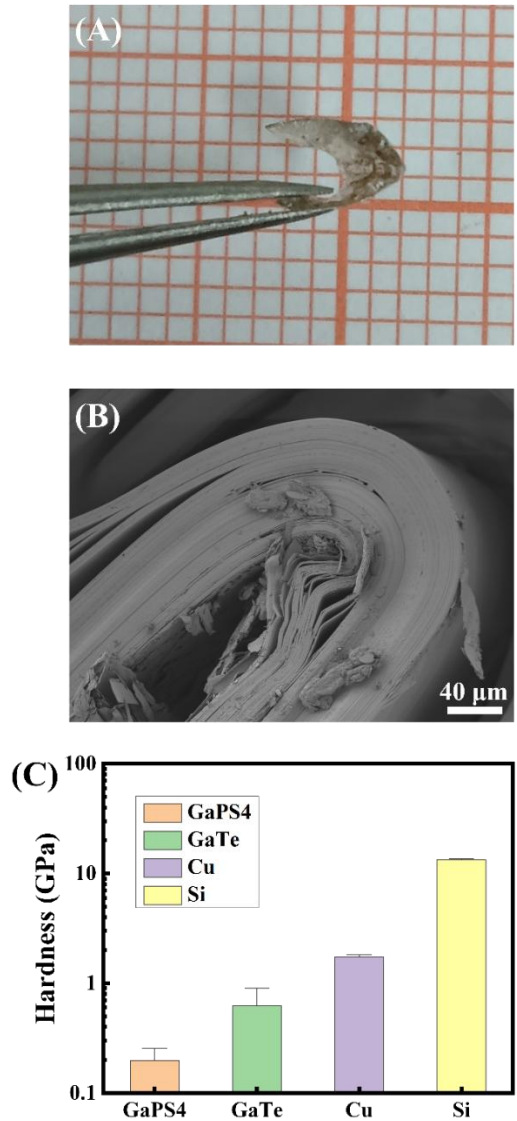

**Figure 2. Deformability of the GaPS₄ single crystal.** (A) Synthetic GaPS$_4$ single crystals are morphed into "U" shapes without breaking. Each small grid represents 1 mm. (B) SEM Morphology of Folded GaPS$_4$ Single Crystals. (C) The hardness of GaPS$_4$ is compared with that of silicon (a classic brittle material), polycrystalline copper (a ductile metal), and GaTe (a gallium-based ductile material). The data are derived from nanoindentation tests.

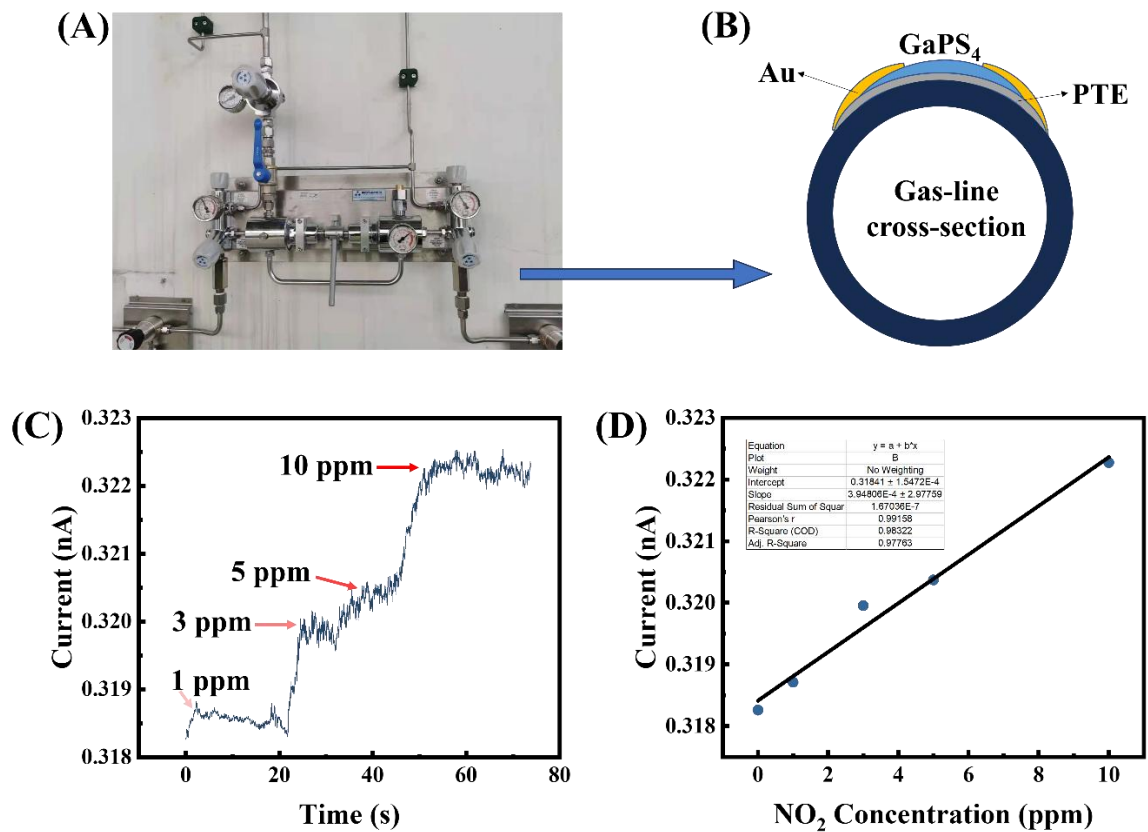

**Figure 3. Conformal Monitoring of Gas-lines with GaPS$_4$ Sensors.** (A) Photograph of the gas-line system in a laboratory setting. (B) Schematic illustration of the GaPS$_4$ gas sensor and its conformal integration onto a gas-line. (C) Current response of the GaPS$_4$ sensor upon exposure to 1, 3, 5, and 10 ppm of NO$_2$. (D) Sensor response current as a function of NO$_2$ concentration.